\documentclass[preprint,aps,pra,amsmath,amssymb,showpacs,superscriptaddress]{revtex4-1}
\usepackage{graphicx}
\usepackage{dcolumn}
\usepackage{bm}
\usepackage{color}
\begin{document}

\title{High-order harmonic generation from 2D periodic potentials in circularly and bichromatic circularly polarized laser fields}

\author{Guang-Rui Jia}\affiliation{College of Physics and Materials Science, Henan Normal University, Xinxiang 453007, China}
\affiliation{State Key Laboratory of Magnetic Resonance and Atomic and Molecular Physics, Wuhan Institute of Physics and Mathematics, Chinese Academy of Sciences, Wuhan 430071, China}
\author{Xin-Qiang Wang}\affiliation{State Key Laboratory of Magnetic Resonance and Atomic and Molecular Physics, Wuhan Institute of Physics and Mathematics, Chinese Academy of Sciences, Wuhan 430071, China}\affiliation{University of Chinese Academy of Sciences, Beijing 100049, China}
\author{Tao-Yuan Du }\affiliation{State Key Laboratory of Magnetic Resonance and Atomic and Molecular Physics, Wuhan Institute of Physics and Mathematics, Chinese Academy of Sciences, Wuhan 430071, China}\affiliation{University of Chinese Academy of Sciences, Beijing 100049, China}
\author{Xiao-Huan Huang}\affiliation{Hubei Key Laboratory of Pollutant Analysis and Reuse Technology, College of Chemistry and Chemical Engineering, Hubei Normal University, Huangshi 435002, China}
\author{Xue-Bin Bian}\email{xuebin.bian@wipm.ac.cn}
\affiliation{State Key Laboratory of Magnetic Resonance and Atomic and Molecular Physics, Wuhan Institute of Physics and Mathematics, Chinese Academy of Sciences, Wuhan 430071, China} 

\begin{abstract}
We studied the high-order harmonic generation (HHG) from 2D solid materials in circularly and bichromatic circularly polarized laser fields numerically by simulating the dynamics of single-active-electron processes in 2D periodic potentials. Contrary to the absence of HHG in the atomic case, circular HHGs below the bandgap with different helicities are produced from intraband transitions in solids with $C_4$ symmetry driven by circularly polarized lasers. Harmonics above the bandgap are elliptically polarized due to the interband transitions. High-order elliptically polarized harmonics can be generated efficiently by both co-rotating and counter-rotating bicircular mid-infrared lasers. The cutoff energy, ellipticity, phase, and intensity of the harmonics can be tuned by the control of the relative phase difference between the 1$\omega$ and 2$\omega$ fields in bicircularly polarized lasers, which can be utilized to image the structure of solids. 
\pacs{42.65.Ky, 42.65.Re, 72.20.Ht}

\end{abstract}
\maketitle
\section{Introduction}\label{I}
High-order harmonic generation (HHG) from atomic and molecular gases has been studied extensively \cite{Krausz,Peng}. It has been utilized to generate attosecond laser pulses. The mechanism is well described by the three-step recollision model \cite{Corkum}. Recently, more attention has been attracted to the HHG from solids \cite{Ghimire,Lee,Huttner,Yu,Liucandong,Liluning,Schubert,Ndabashimiye,You2} with the development of long-wavelength lasers. Solid HHG demonstrates novel characters different from the HHG from gases. For example, linear cutoff energy dependence on the amplitude of the laser field \cite{Ghimire}, multi-plateau structure in the HHG spectra \cite{Ndabashimiye}, and different laser ellipticity dependence \cite{Ghimire,Liucandong,Liluning,Nicolas,You,Yoshikawa,Tamaya,Saito}. However, mechanisms of HHG from solids are still under debate. Inter- and intra-band transition models \cite{Pronin1,Pronin2,Guan,Du2,Vampa,Du3} are proposed. Three-step model in coordinate space \cite{Vampa} and step-by-step model in vector $k$ space \cite{Du1,Jia,Wu2,Ikemachi} are investigated. The drawback of the solid HHG is the low damage threshold of solid materials. Many efforts have been undertaken to enhance the yield of HHG. For example, two-color laser fields \cite{Vampa3,Li,LiuXi} and plasmon-enhanced inhomogeneous laser fields \cite{Du2,Vampa2,Joel,Han} are used to manipulate the HHG process, especially for the enhancement of the second plateau of the HHG spectra \cite{Du2}.

In circularly polarized laser fields, HHGs are absent in the atomic systems due to the non-recollision spin motions of electrons in the classical picture \cite{Corkum} and forbidden transition in the selection rules in the quantum picture. However, HHG occurs in molecular systems in circularly polarized lasers because of additional recollision centers \cite{Yuan1}. In the solid systems, especially for 2D materials \cite{Liu,Yoshikawa,Saito}, because of the multi-center periodic potential wells and delocalization of the wave packets, HHGs can be efficiently produced in circularly polarized driving laser fields with the polarization plane in the same plane of 2D solids \cite{Liluning,Saito}. The intensity of HHG as a function of the driving laser ellipticity has been studied recently \cite{Ghimire,Nicolas,You,Yoshikawa,Tamaya}. In this work, we study the ellipticity of HHG from solids driven by the external circular and bicircular laser fields. 
   
In bicircular laser fields \cite{Zuo,Eichmann,Long,Milosevic1,Milosevic2}, especially for the counter-rotating case, circular HHGs with different helicities can be efficiently generated in atomic and molecular systems \cite{Zuo,Yuan2,Mauger}. This has been experimentally demonstrated recently \cite{Kfir}. It can also be used to illuminate chirality \cite{Milosevic3} and  molecular symmetries \cite{Bay,Zhu,Reich}. It is also possible to generate isolated elliptical and circular attosecond laser pulses \cite{Milosevic4,Lukas,Yuan3,Xia}. However, the HHG from solids in bicircular laser fields is rarely investigated. In this work, we study the electron dynamics in 2D periodic potentials \cite{Hawkins} to simulate the HHG process in solids. Atomic units are used throughout. 

\section{Numerical results by solving the time-dependent Schr\"odinger equation}\label{III}
\begin{figure}
	\centering\includegraphics[width=9 cm,height=8 cm]{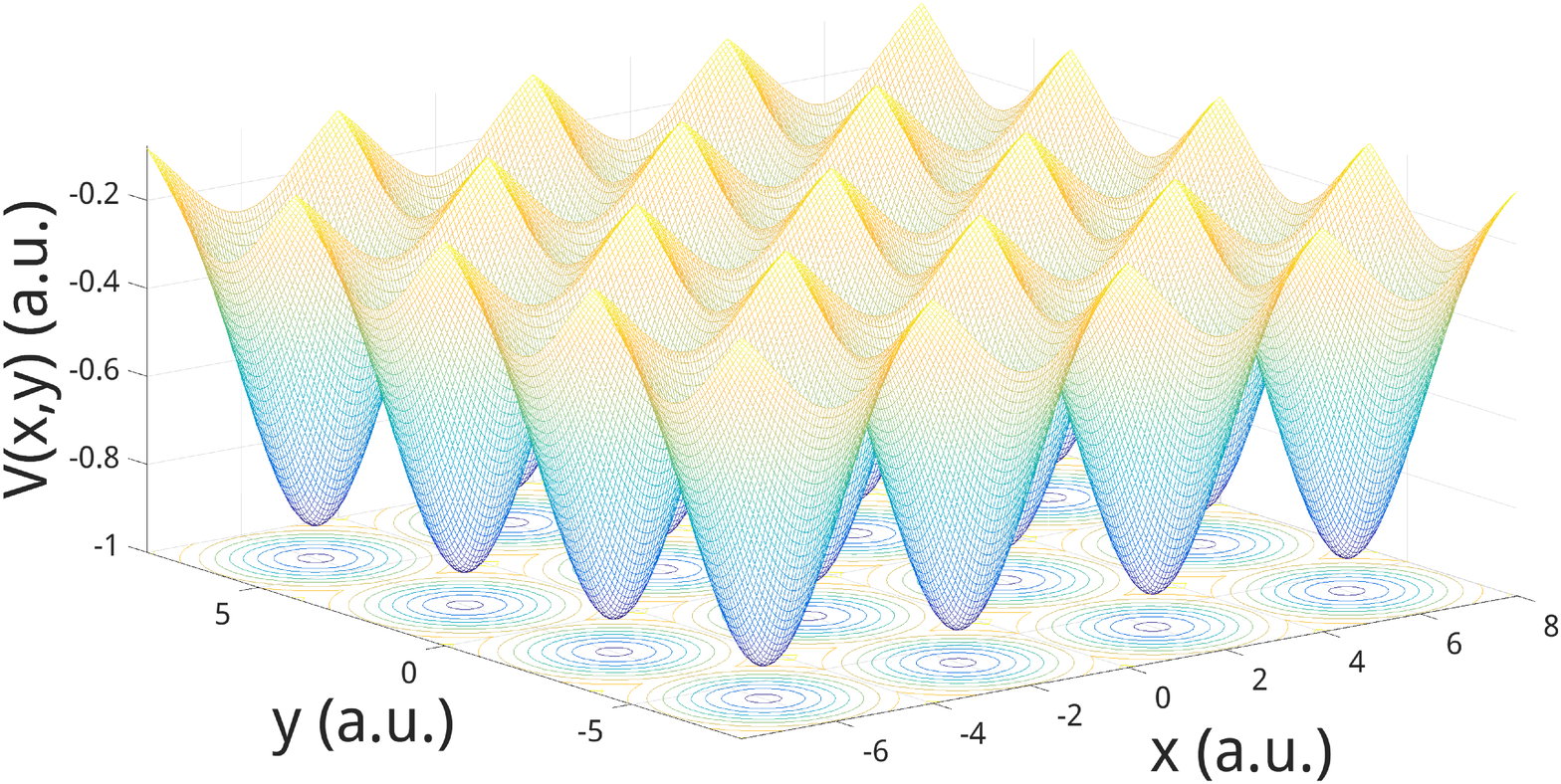}
	\caption{Illustration of the 2D periodic potentials in the range of $x,y \in[-8, 8]$ a.u. The projection of the potentials is in the bottom, showing the $C_4$ symmetry of the system.}\label{Fig1}
\end{figure}
In the single-active-electron approximation, the time-dependent Schr\"odinger equation(TDSE) can be written as
\begin{equation}
         i\dfrac{\partial \Psi(x,y,t)}{\partial t}=(\dfrac{p_x^2}{2}+\dfrac{p_y^2}{2}+V(x,y)+xE_x(t)+yE_y(t))\Psi(x,y,t)
         \end{equation}
where $E_x(t)$ and $E_y(t)$ are the $x$ and $y$ components of the laser fields. $V(x,y)$ are two-dimensional periodic Gaussian potentials \cite{Pavelich}. The form for one unit cell is
\begin{equation}
         V_c(x,y)=-V_0 \exp\left\lbrace -\left[\alpha_x\dfrac{(x-x_0)^2}{a^2}+\alpha_y\dfrac{(y-y_0)^2}{a^2} \right]  \right\rbrace   
         \end{equation}
where $V_0$ represents the maximum depth of the potential well, $a$ is the length of the unit cell, ($x_0$, $y_0$) are the coordinates of the center of the potential well. In this work, $a=4$ a.u., $V_0=3\pi^2/2a^2$, $\alpha_x=\alpha_y=6.5$. As illustrated in Fig. \ref{Fig1}, this 2D system has $C_4$ symmetry. The electrons in the valence band are mainly distributed in each potential well, while the the electrons in the conduction bands are more delocalized.

\begin{figure*}
	\centering\includegraphics[width=12 cm,height=6 cm]{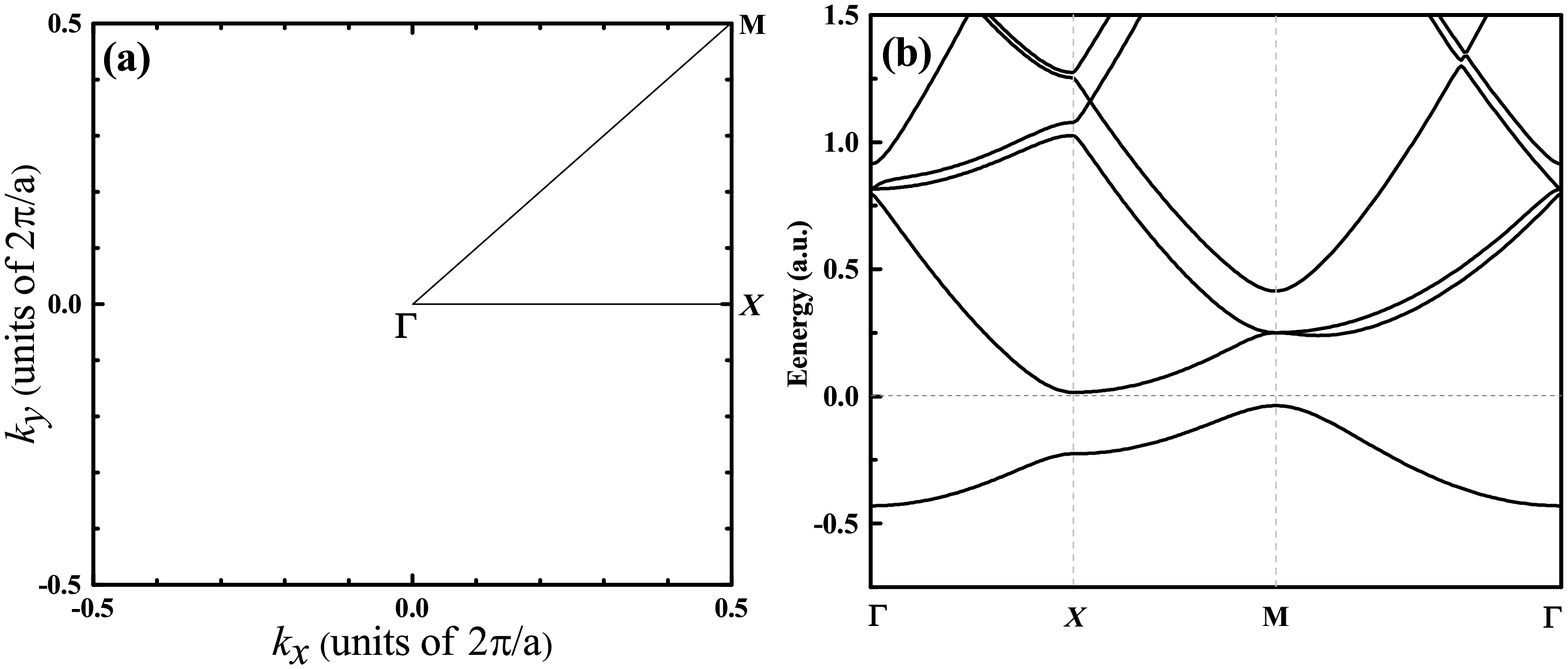}
	\caption{Brillouin zone and the energy band structure of the 2D potentials. $\Gamma$, X and M represent the high symmetry points.}\label{Fig2}
\end{figure*}

Based on Bloch's theorem \cite{Du1}, the band structure of the system can be obtained by diagonalization of the field-free Hamiltonian matrix in reduced Brillouin zone. The results are shown in Fig. \ref{Fig2}. $\Gamma$, X and M represent the high symmetry points illustrated in Fig. \ref{Fig2}(a).

We solve the TDSE in the real space in the range $x,y\in[-200,200]$ a.u. The wave functions are expanded by B-splines:
\begin{equation}
         \Psi(x,y,t)=\sum_{i,j}C_{i,j}(t)B_i(x)B_j(y)
         \end{equation}

 The time-dependent wavefunction is obtained by using the Crank-Nicholson method \cite{Bianc,Guan}. An absorbing function is used at the boundary to remove artificial reflection. The laser-induced currents along $x$ and $y$ axes are:

\begin{eqnarray}
j_x(t) &=& \langle\Psi(x,y,t)\vert \hat{p}_x\vert\Psi(x,y,t)\rangle, \nonumber \\
j_y(t) &=& \langle\Psi(x,y,t)\vert \hat{p}_y\vert\Psi(x,y,t)\rangle. 
\end{eqnarray}

The HHG spectra are calculated by Fourier transforms of the above currents. The obtained complex $xy$ components $\sqrt{P_x(\Omega)}\exp\left[i\Phi_x(\Omega) \right]$ and $\sqrt{P_y(\Omega)}\exp\left[i\Phi_y(\Omega) \right]$ of harmonic photon with energy $\Omega$ can be used to extract its ellipticity $\varepsilon$ and phase $\delta$. They are defined as \cite{Yuan1,Born}:
\begin{equation}
\varepsilon = \tan \chi, 
\end{equation}
where
\begin{eqnarray}\label{Eq6}
\sin(2\chi) &=& \sin(2\gamma)\sin \delta, \nonumber \\
\tan \gamma &=& \sqrt{P_y(\Omega)/P_x(\Omega)}, \nonumber \\
\delta &=& \Phi_y(\Omega)-\Phi_x(\Omega).
\end{eqnarray}
The values of $\varepsilon$ and $\delta$ are in the range of [-1, 1] and [0, 2$\pi$], respectively. $\varepsilon>0$ represents right polarization, while $\varepsilon<0$ stands for left polarization.             
\section{High-order harmonic generation in circularly polarized laser fields}
\begin{figure}
\centering\includegraphics[width=12 cm,height=7 cm]{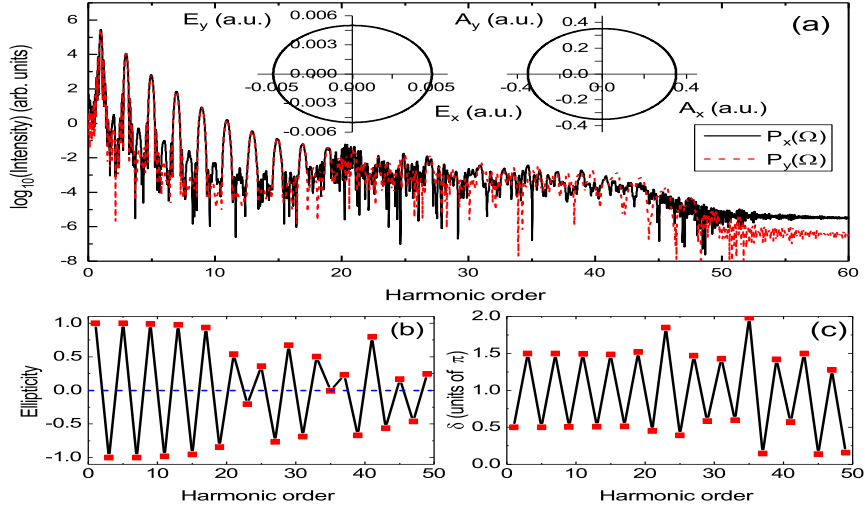}
\caption{HHG spectra from 2D periodic potentials in circularly polarized laser fields. (a) HHG spectra along the $x$ and $y$ axes. The inset is the Lissajous curves of the electric field and vector potential for one optical cycle of the laser field. (b) The ellipticity of each harmonic. (c) Phase difference between the $x$ and $y$ components of harmonic emission. The intensity of the laser is $I=8.77\times10^{11}$W/cm$^2$, the wavelength $\lambda=3.2$ $\mu$m. The initial state in the TDSE calculation is the state on top of valence band.}\label{Fig3}
\end{figure}

The circularly polarized laser field is written in the following form:
\begin{eqnarray}\label{Eq7}
E_x(t) &=& E_0f(t)\sin(\omega t+\phi), \nonumber \\
E_y(t) &=& E_0f(t)\cos(\omega t+\phi). 
\end{eqnarray}
where the pulse envelope is
\begin{equation}\label{Eq8}
         f(t)=\cos^2(\pi(t-\tau/2)/\tau), 
\end{equation}        
$\tau$ is the total pulse duration, which is set to be 10 cycles in this work. $E_0=0.005$ a.u. The wavelength is $\lambda=$3.2 $\mu$m. 

At first, the initial state in our TDSE evolution is the single state on top of the valence band, i.e., the state at M point in Fig. \ref{Fig2}(b). The HHG spectra are presented in Fig. \ref{Fig3}.
\begin{figure}
\centering\includegraphics[width=12 cm,height=7 cm]{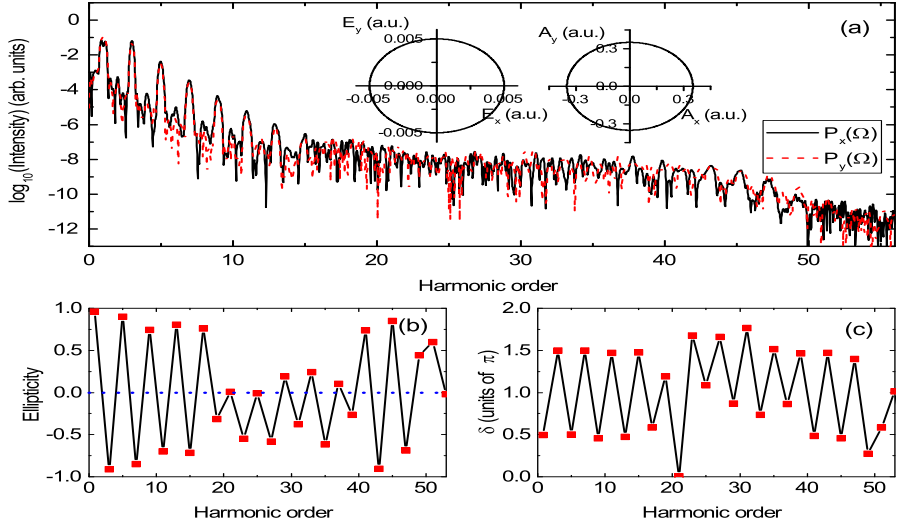}
\caption{Same as Fig. \ref{Fig3} except that the initial state in the TDSE calculation is a superposition of all valence states.}\label{Fig4}
\end{figure}

The ground state of atoms is localized to the atomic core. Without recollision of ionized electron to the parent ion in the circularly polarized driving lasers, HHG can not be produced no matter how intense the lasers are. Different from the absence of HHG from atoms in circularly polarized driving laser fields, clear HHG signals with long plateau are generated in our simulations. Due to the multiple potential wells, the electron states in both the valence band and conduction band are delocalized. The transition probability between them is high even in the circularly polarized laser fields. In Fig. \ref{Fig3}, strong odd harmonics are dominant. One can find that circularly polarized odd harmonics with order less than 19 are generated in Fig. \ref{Fig3}(b). This energy is close to the minimum band gap between the valance band and the first conduction band: $E_c-E_v=0.25$ a.u. This is in the perturbation regime. The harmonics are mainly from the intra-band transitions from previous studies \cite{Guan,Du3}. They reflect the global motion of the electron wavepackets along the conduction band \cite{Du3}. The motion of electrons closely follows the spinning electric field from the circularly polarized driving lasers. Consequently, circular HHGs are produced. Different from the transition selection rules in the atomic case, circular harmonics with different helicities are allowed to be produced from solids with different symmetries \cite{Tang}. The selection rules for 2D solids with $C_4$ symmetry in circular driving lasers in \cite{Tang} explain well why the helicity of HHG changes alternately in Fig. \ref{Fig3} (b). For harmonics with order $N>20$, they are generated from the inter-band transitions with a plateau structure in the non-perturbation regime \cite{Guan,Du3}. The harmonics reflect the local instantaneous polarization between the conduction band and the valence band of the system \cite{Du3}. The inter-band transitions do not closely follow the driving circular fields. Elliptically polarized harmonics are presented instead. One may also notice the weak even-order harmonics with $N<20$ in Fig. \ref{Fig3}, which are forbidden in the transition selection rules in \cite{Tang}. However, the rules are based on group theories, which are applicable in global intraband oscillations rather than local interband polarization. To check if it is from the asymmetry introduced by the finite laser pulses, we have increased the pulse duration to 20 cycles. The weak even harmonics remain, which suggests that they are from the influence of interband transitions, similar to the case of interband-transition-induced even-order HHG in linearly polarized laser fields \cite{Jia}. The cutoff energy is determined by the maximum energy band gap between the first conduction band and the valence band at all possible wave vector $\vec{k}(t)$ from our proposed model \cite{Du1} in the momentum space. $\vec{k}(t)=\vec{k}(0)+\vec{A}(t)$, where $\vec{A}(t)$ is the vector potential of the laser fields and $\lvert k(0)_x\rvert=\lvert k(0)_y\rvert=\pi/a$ (M point). From Fig. \ref{Fig2}, the value of the maximum band gap between the valence band and the first conduction band is around 41$\omega$, agreeing well with the cutoff energy in Fig. \ref{Fig3}, suggesting that the HHG model for interband transition \cite{Du1} is valid even in circularly polarized laser fields.

The bandgap along $X-M$ is also small as illustrated in Fig. \ref{Fig2}. To include the contribution of all valence states, we use a normalized superposition of all valence states as the initial state to do the TDSE simulations. The results are presented in Fig. \ref{Fig4}. One may find that the harmonics with order below 20 do not have perfect left or right circular polarization as shown in Fig. \ref{Fig3}, but elliptical polarization. It is due to the coupling of different valence states. The similar thing occurs in the HHG from atoms in bi-circular lasers by involving the Rydberg states \cite{Ivanov}. Dynamic symmetry breaking of the system introduced by the superposition state will change the polariztion of HHG. It may also explain why the experimentally measured ellipticity of HHG  is not exactly 1 \cite{Saito}. In all the following calculations, we use the above superpostion state as the initial state for the TDSE simulations.

\section{High-order harmonic generation in co-rotating two-color circularly polarized laser fields}
\begin{figure}
\centering\includegraphics[width=11 cm,height=8 cm]{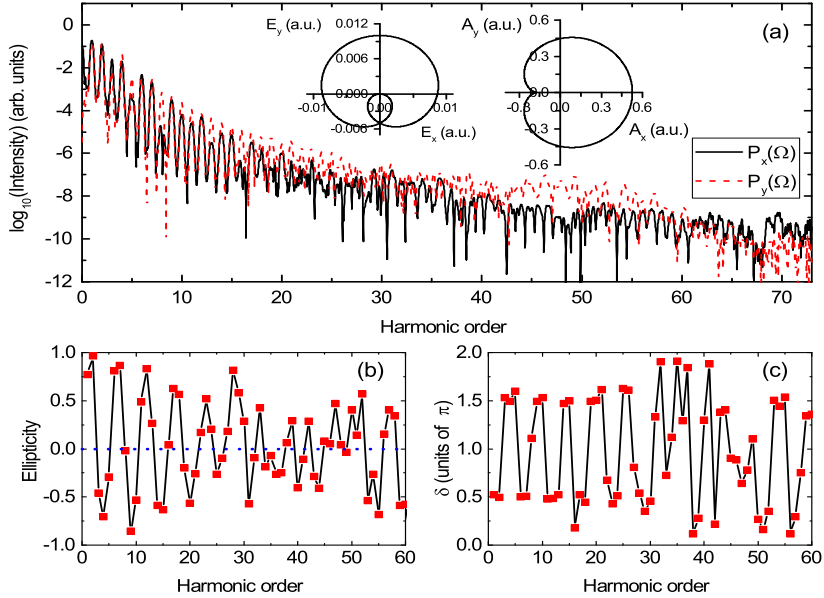}
\caption{HHG spectra from 2D periodic potentials in co-rotating two-color circularly polarized laser fields. (a) HHG spectra along the $x$ and $y$ axes. The inset is the Lissajous curves of the electric field and vector potential for one optical cycle of the fundamental field. The phase $\phi=0$. (b) The ellipticity of each harmonic. (c) Phase difference between the $x$ and $y$ components of harmonic emission. The intensity of the laser is $I=8.77\times10^{11}$W/cm$^2$, the wavelength of the fundamental laser is $\lambda=3.2$ $\mu$m.}\label{Fig5}
\end{figure}
\begin{figure}
\centering\includegraphics[width=12 cm,height=9 cm]{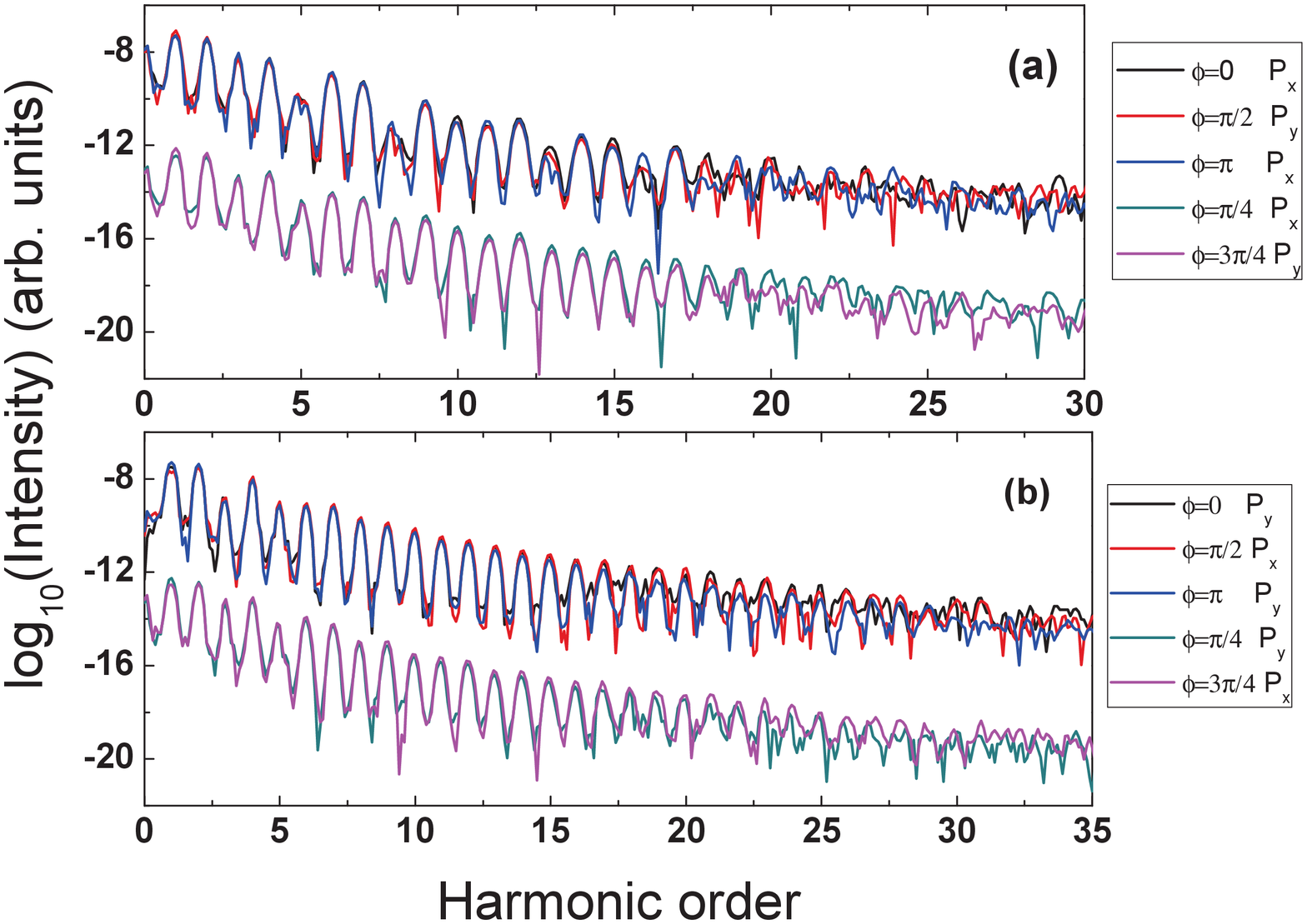}
\caption{HHG spectra in co-rotating two-color circular lasers with different phases. The parameters of the laser field are the same as those in Fig. \ref{Fig5} except for different phases marked in the figure. The spectra are shifted for clarity.}\label{Fig6}
\end{figure}
\begin{figure}
	\centering\includegraphics[width=9 cm,height=5 cm]{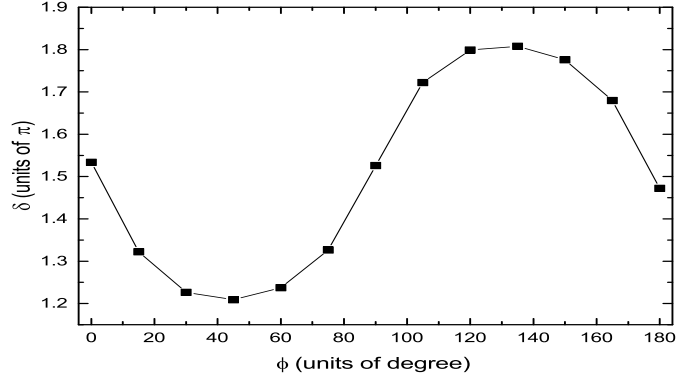}
	\caption{Phase difference of the third harmonic as a function of the phase in co-rotating lasers.}\label{Fig7}
\end{figure}
Two-color linearly polarized lasers have been shown to be efficient tools to generation high harmonics \cite{Vampa3,Zaks}. In this work, we investigate the role of two-color circularly polarized lasers in HHG. The co-rotating two-color $1\omega$ and $2\omega$ laser fields are written as
\begin{eqnarray}\label{Eq9}
E_x(t) &=& E_0f(t)(\sin(\omega t)+\sin(2\omega t+\phi)), \nonumber \\
E_y(t) &=& E_0f(t)(\cos(\omega t)+\cos(2\omega t+\phi)). 
\end{eqnarray}
The intensities of the two lasers are set to be the same. The pulse shape and the durations are the same as those in Eq. (\ref{Eq7}). The phase $\phi$ will rotate the electric fields and vector potentials of the laser fields. The rotation angle in the Lissajous curve is equal to this phase. It plays no roles in the HHG from atoms due to the isotropy, while it plays important roles in HHG from solids. The HHG spectra with different phases $\phi$ are illustrated in Figs. \ref{Fig5} and \ref{Fig6}. For the HHG from atoms, this co-rotating scheme is not efficient \cite{Eichmann} since the recollision probability is still small. However, a long plateau with clear cutoff beyond the minimum band gap can be observed from the HHG spectra in the above figures. This scheme of co-rotating two-color circular lasers is also an efficient way to generate HHGs. Different from the case of circular driving lasers, even-order harmonics are obviously generated and their intensities are comparable to those of their neighboring odd harmonics. One may find that the harmonics with order $N<19$ below the minimum band gap are not circularly polarized either, even though some of their ellipticities are close to 1. We also use the single state on top of the valence band to do TDSE simulations (not shown here), no circular harmonics are observed. It is mainly attributed to the different laser strengths along $x$ and $y$ axes as illustrated in the inset of Fig. \ref{Fig5}.

In the multiphoton picture \cite{Eichmann,Vampa3,Zaks}, harmonic with order $N$ may come from different combinations of $N_{1}$ $\omega$ and $N_{2}$ $2\omega$ photons ($N_{1}+2N_{2}=N$) with different strengths and phases. The interference of these channels makes the phases of the harmonics complex. One may find that the harmonic with order 8 shown in Fig. \ref{Fig5} is approximately linearly polarized.

Orientation dependent harmonic generation in linearly polarized pulses is a good way of probing solid structure \cite{You,Wu3,Nagler}. It can be extended to bicircular laser fields. To image the structure of the solid material, harmonic intensities along $x$ and $y$ axes with different phases $\phi$ in the driving lasers are presented in Fig. \ref{Fig6}. The intensities of harmonics below the bandgap as a function of $\phi$ exhibit a period of $\pi/2$, reflecting well the $C_4$ symmetry of the system. It can be a useful tool for probing the structure of solid target. However, the harmonics contributed by the interband polarization above the bandgap with order $N>19$ loose the periodicity since they are very sensitive to the laser fields, especially for short pulses \cite{Guan}. The phase in Eq. (\ref{Eq9}) is equivalent to a delay of the two-color pulses. This delay between the two drivers will amplify the effect of finite pulse duration on the instantaneous interband transitions. The harmonic above the bandgap can not be used as an imaging tool. We also calculated the phase difference $\delta$ in Eq. (\ref{Eq6}) of the third harmonic as a function of the phase $\phi$ of the lasers in Eq. (\ref{Eq9}), which is shown in Fig. \ref{Fig7}. One may find that it shows a half period of $\pi/2$ and sensitivity to the external laser fields, which can be used to image the symmetry group of the solid structure efficiently. 

From Fig. \ref{Fig2}, the band structure of the system is anisotropic. As a result, the phase $\phi$ in the co-rotating driving lasers can be used to control the HHG process, and thus the properties of the HHG signals, such as the cutoff energy, the relative intensity, and phase difference of HHG along $x$ and $y$ directions.
\begin{figure}
	\centering\includegraphics[width=11 cm,height=8 cm]{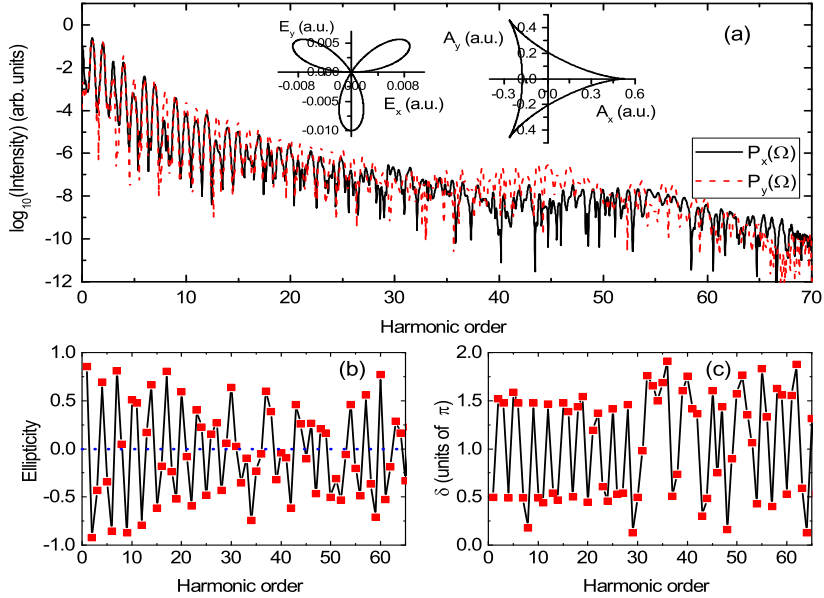}
	\caption{HHG spectra from 2D periodic potentials in counter-rotating two-color circularly polarized laser fields. (a) HHG spectra along the $x$ and $y$ axes. The inset is the Lissajous curves of the electric field and vector potential for one optical cycle of the fundamental field. The phase $\phi=0$. (b) The ellipticity of each harmonic. (c) Phase difference between the $x$ and $y$ components. The intensity of the lasers is $I=8.77\times10^{11}$W/cm$^2$, the wavelength of the fundamental laser is $\lambda=3.2$ $\mu$m.}\label{Fig8}
\end{figure}
\section{High-order harmonic generation in counter-rotating two-color circularly polarized laser fields}

\begin{figure}
\centering\includegraphics[width=12 cm,height=10 cm]{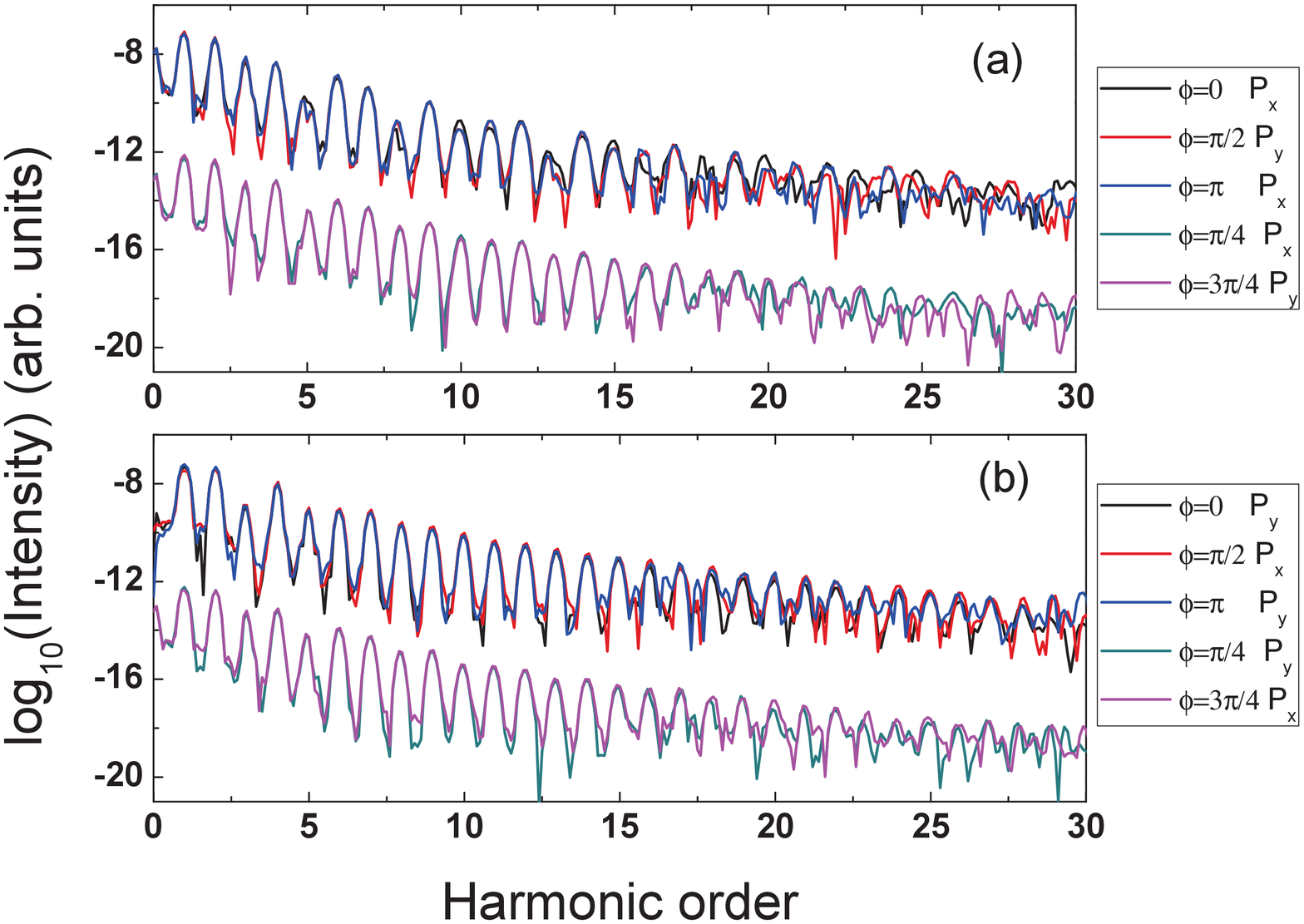}
\caption{HHG spectra in counter-rotating two-color circular lasers with different phases. The parameters of the laser field are the same as those in Fig. \ref{Fig8} except for different phases marked in the figure. The spectra are shifted for clarity.}\label{Fig9}
\end{figure}

\begin{figure}
\centering\includegraphics[width=9 cm,height=7 cm]{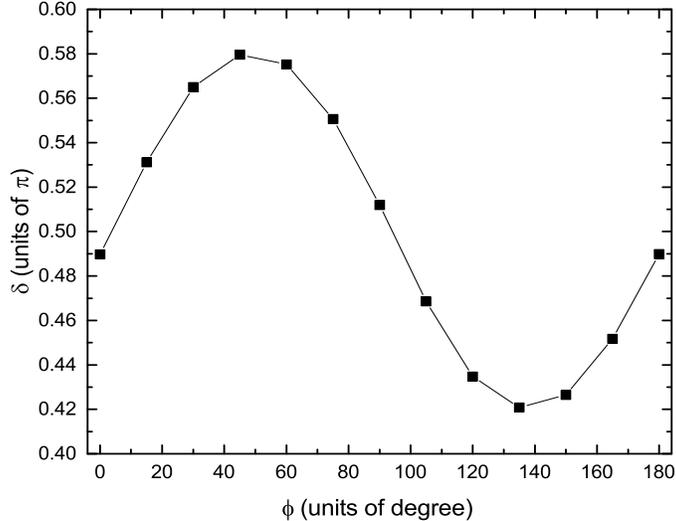}
\caption{Phase difference of the fourth harmonic as a function of the phase in counter-rotating lasers.}\label{Fig10}
\end{figure}

The counter-rotating bicircular $1\omega$ and $2\omega$ laser fields with amplitude ratio 1:1 are written as
\begin{eqnarray}\label{Eq10}
E_x(t) &=& E_0f(t)(\sin(\omega t)+\sin(2\omega t+\phi)), \nonumber \\
E_y(t) &=& E_0f(t)(\cos(\omega t)-\cos(2\omega t+\phi)). 
\end{eqnarray}
The calculated HHG spectra with different $\phi$ are shown in Figs. \ref{Fig8} and \ref{Fig9}. The rotation angle in the Lissajous curve is also the same as $\phi$.

Circular HHGs with different helicities are generated from atoms in bicircular counter-rotating laser fields \cite{Milosevic1}. The harmonics with order $3N$ are forbidden from the transition selection rule in atomic systems. However, in Fig. \ref{Fig8}, no perfect circular HHGs are produced in the case of solids. Some harmonics close to linear polarization are also generated in counter-rotating bicircular laser fields. These features suggest the different mechanisms of HHG from gases and solids. The harmonic order 3$N$ is not restricted by transition selection rules in this $C_4$-symmetry potential wells. For the case of HHG from atoms, co-rotating bicircular lasers are less efficient to generate harmonics compared to the counter-rotating bicircular lasers due to the necessary condition of recollision. However, the efficiencies for generating HHG from 2D solids are comparable in the co-rotating and counter-rotating bicircular laser fields since the electron wave functions in both the valence band and conduction band are delocalized and the condition of electrons returning to its original position is unnecessary.

The phase $\phi$ can be used to tune the ellipticity, the cutoff energies, relative intensity, and phase difference $\delta$ of the $xy$ components of the HHGs. It can also be used to image the structure of solids. As shown in Fig. \ref{Fig9}, the harmonic intensities along $x$ and $y$ directions below the bandgap demonstrate a period of $\pi/2$. However, to retrieve the structure of an unknown target, one should fit the intensity of a range of harmonics to achieve high sensitivity. We also calculated the phase $\delta$ of the third and fourth harmonics as a function of the phase $\phi$ of the lasers in Eq. (\ref{Eq10}). The phase of the third harmonic demonstrates a similar -sine trend in co-rotating lasers illustrated in Fig. \ref{Fig7}, while the phase of the fourth harmonic shows a sine trend in Fig. \ref{Fig10}. One may find that it also demonstrates a half period of $\pi/2$ and sensitivity to the change of the external laser fields, which can be used to image the solid structure efficiently. 

The amplitude ratio of the bicircular lasers will also affect the HHG signals. The combinations of lasers are not restricted to the fundamental and its second harmonic (1$\omega$+2$\omega$), it can be generalized to any $n\omega+m\omega$ bicircular field to control HHG in soilds. There are too many ways to combine these parameters to control the HHG processes. Their effects will not be discussed in this work.
\section{Summary}\label{IV}

 In conclusion, we have studied the HHG from 2D solids in circular and bicircular co-rotating and counter-rotating laser fields. Different from the HHG from atoms, circular HHGs from intraband transitions can be generated in solids by circular driving lasers by using a pure initial valence state. Contributions from different valence states will change the polarization of harmonics. The efficiencies to generate harmonics from co-rotating and counter-rotating bicircular lasers are comparable. Elliptically and linearly polarized harmonics can be generated. The ellipticity, the relative intensity and phase difference between the $xy$ harmonic components, and the cutoff energy can be controlled by the phase of the bicircular driving lasers. This phase dependence reflects the spacial distribution and the energy band structure of the solid targets, which can be used as an imaging tool.

\section{ACKNOWLEDGMENTS}\label{V}
We thank XuanYang Lai and HongChuan Du for many very helpful discussions. This work is supported by the National Natural Science Foundation of China(NSFC) (11561121002, 21501055, 11404376, 11674363, 61377109), Youth Science Foundation of Henan Normal University (2015QK03), Start-up Foundation for Doctors of Henan Normal University (QD15217).

\end{document}